\documentclass[aps,prl,reprint,superscriptaddress]{revtex4-2}

\usepackage{graphicx}
\usepackage{epstopdf}
\usepackage{bm}
\usepackage{hyperref}
\usepackage{color}
\usepackage{easyReview}
\usepackage{latexsym}

\usepackage{changes}

\begin{document}

\title{A high-performance surface acoustic wave sensing technique}

\author{Mengmeng Wu}
\affiliation{International Center for Quantum Materials, 
	Peking University, Haidian, Beijing, 100871, China}
\author{Xiao Liu}
\affiliation{International Center for Quantum Materials, 
	Peking University, Haidian, Beijing, 100871, China}
\author{Renfei Wang}
\affiliation{International Center for Quantum Materials, 
	Peking University, Haidian, Beijing, 100871, China}
\author{Xi Lin} 
\email{xilin@pku.edu.cn}
\affiliation{International Center for Quantum Materials, 
	Peking University, Haidian, Beijing, 100871, China}
\affiliation{Interdisciplinary Institute of Light-Element Quantum Materials and Research Center for Light-Element Advanced Materials, Peking University, Haidian, Beijing, 100871, China}
\author{Yang Liu} 
\email{liuyang02@pku.edu.cn}
\affiliation{International Center for Quantum Materials, 
	Peking University, Haidian, Beijing, 100871, China}

\date{\today} 

\begin{abstract}

  We present a superheterodyne-scheme demodulation system which can
  detect the amplitude and phaseshift of weak radio-frequency
  signals with extraordinarily high stability and resolution. As a
  demonstration, we introduce a process to measure the velocity of the
  surface acoustic wave using a delay-line device from 30 K to room
  temperature, which can resolve $\textless$ 0.1ppm velocity
  shift. Furthermore, we investigate the possibility of using this
  surface acoustic wave device as a calibration-free, high sensitivity
  and fast response thermometer.

\end{abstract}
                      
\maketitle

The surface acoustic wave (SAW) is a special type of acoustic mode which
 propagates along the surface of a solid within a length scale of about
its wavelength \cite{white1965direct}. Environmental perturbations
directly affect the propagation velocity $v$ and the insertion loss
$\mit\Gamma$ of SAW. Therefore, SAW is broadly used as sensors in many
applications with its unique advantages such as compact size, low cost
and ease of fabrication \cite{white1985surface, benes1995sensors,
  liu2016surface}. In the past few decades, SAW sensors have been
developed for various applications such as temperature
\cite{hauden1981temperature, viens1990highly, neumeister1990saw},
pressure \cite{scherr1996quartz,jungwirth2002micromechanical}, strain
\cite{fu2014stable}, or for biosensing applications
\cite{voiculescu2012acoustic, huang2021surface}. SAW is also a very
useful current-free technique in the study of quantum phenomena. It has been used to investigate the property of two
dimensional system and prevail the nature of fragile quantum states
\cite{Wixforth.PRL.1986, Paalanen.PRB.1992,Willett.PRL.1993,
  Willett.PRL.2002, Friess.PRL.2018,Friess.PRL.2020,
  Friess.Nature.2017, Drichko.PRB.2011,Drichko.PRB.2016}.
	
In piezoelectric materials, the SAW traveling along the surface
converts electric fields to mechanical motion and vice versa. It can
be generated and detected by interdigital transducers (IDTs) when the
frequency of applied/SAW-induced oscillating voltage matches the
resonance condition $f_r=v/\lambda_r$, where $\lambda_r$ is the period of
the IDT. The SAW velocity $v$ changes when the external physical
properties such as temperature, pressure, etc. vary. SAW sensor
devices can be categorized into two types. The first type is a resonator
with the central IDT between two reflecting gratings that are totally reflecting at the desired resonance frequency $f_r$ \cite{benes1995sensors, viens1990highly}. It is included into the
feedback loop of an oscillator for frequency readout.  The second type
has a delay-line structure consisting of spatially separated emitter and
receiver IDTs. The variation of the sensing region between the IDTs
alters the phase $\mit\Phi$ of the received signal
\cite{Wixforth.PRL.1986, Paalanen.PRB.1992,Willett.PRL.1993,
  Willett.PRL.2002, Friess.PRL.2018,Friess.PRL.2020,
  Friess.Nature.2017, Drichko.PRB.2011,Drichko.PRB.2016} or the
resonance frequency $f_r$ of the device
\cite{neumeister1990saw,powlowski2019temperature,fu2014stable}. The
resonator approach suffers from intrinsic electronics noise, parasitic
oscillations and saturation of the oscillator
\cite{durdaut2018oscillator}. In practice, the delay-line device is
preferred for its better precision and reliability.
 
In this work, we introduce a superheterodyne-scheme radio frequency
(RF) lock-in technique which can measure the phase ($\mit \Phi$) of a -90
dBm signal with sub-mrad resolution. We then demonstrate an
application of SAW delay-line devices for temperature sensing
and calibrate the SAW traveling time $\tau$ from $\sim$30 K to room
temperature. At $\tau\simeq$ 1 $\mu$s and $f_{\text{r}}\simeq$ 560
MHz, we can detect $\sim$0.1-ppm of $\frac{\Delta \tau}{\tau}$,
corresponding to $\sim 1$ mK temperature resolution at room
temperature. Our setup with its combination of low excitation and high
sensitivity shows potential to resolve very delicate responses of
fragile quantum phases that are only visible at extremely low
temperatures. In a separate work, we demonstrate such an application
by studying quantum Hall states in a two dimensional system at mK
temperature \cite{wu2023morphing}.
 
\begin{figure}[!htbp]
  \includegraphics[width=0.45\textwidth]{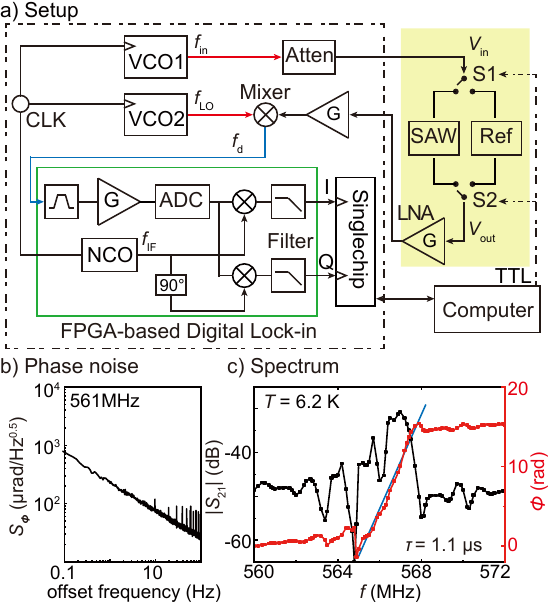}
  \caption{(a) Diagram of superheterodyne quadrature phase
    demodulation system. The FPGA-based digital lock-in module is shown in the green box. (b) Phase noise spectral density measured at
    room temperature. (c) The amplitude and phase of the SAW device
    measured by our setup as a function of frequency.}
  \label{Fig1}
\end{figure}

In order to measure the phase delay of our device, we develop a
heterodyne-scheme lock-in technique that functions from 50 MHz to 1.5
GHz and has an exceptional phase noise performance. Fig. 1(a) shows a
diagram of our setup. We use fractional phase locked-loops (PLL) to
generate two single-frequency signals $V_{\text{in}}$ and
$V_{\text{LO}}$, whose frequencies are locked to the reference clock
frequency $f_{\text{CLK}}$ by $f_{\text{in}}$ (=
$\frac{n}{m}\cdot f_{\text{CLK}}$) and $f_{\text{LO}}$ (=
$\frac{n+k}{m}\cdot f_{\text{CLK}}$) respectively; $n$,$m$ and $k$ are
integer values. $V_{\text{in}}$ is attenuated and sent to the device
which attenuates the amplitude by $\mit \Gamma$ and introduces a phase
delay $\mit \Phi$. The output signal of the device $V_{\text{out}}$
will be magnified by a broadband low noise amplifier (LNA) and then
multiplied with $V_{\text{LO}}$ in the mixer. We then use a filter to
extract the differential frequency component $f_{\text{d}}$ and
digitize it with an audio-frequency A/D converter. We properly choose
the internal Numerically Controlled Oscillator (NCO) module to
generate the intermediate frequency (IF) at
$f_{\text{IF}} = |f_{\text{in}}-f_{\text{LO}}| = \frac{k}{m}\cdot
f_{\text{CLK}}$ and use it as the reference signal for the digital
lock-in-type signal analysis of $V_{\text{d}}$. The amplitude of
$V_{\text{d}}$ is proportional to $V_{\text{out}}$, and its phase
shift $\mit \Phi_{\text{d}}$ from IF is offset from $\mit \Phi$ by a
constant. In particular, we construct a reference channel with fixed
attenuation and phase delay. Two microwave mechanical switches toggle
between the device and reference channels to eliminate the constant
phase offset and compensate for the drift of components such as
amplifiers and mixers \cite{fu2014stable}.

Our system is specifically optimised for the best phase stability and
low-level RF signal, which is required in studying quantum physics
\cite{Zhao.RSI.2022, wu2023morphing}. The superheterodyne-scheme
downconverts the RF signal to intermediate frequency (IF) where
quadrature demodulation is performed, thereby mitigating the 1/f noise
\cite{miller1972high}. We use two fractional phase locked loops
(PLLs) to generate the measurement and reference signals, so that the
phase stability is improved by the internal feedback loop in the PLLs
\cite{woo1973digital}. In contrast to the high speed digitization used in many systems, our intermediate frequency is fixed at
audio-frequency so that a low-noise A/D converter can be used and the high
precision quadrature demodulation algorithm can be implemented.
In addition, the phase noise from the clock-jittering can be
eliminated by carefully matching the clocks used by PLLs, A/D
converters and digital signal processors. Furthermore, the system
features well-designed shielding between its various components and
achieves an isolation of more than 80 dB between two channels. These
precautions effectively reduce background noise and crosstalk
interference.

In typical measurement conditions, the noise is almost always limited
by the phase noise of the PLLs. The measured phase noise spectral
density $S_\phi$(f), shown in Fig. 1(b), is proportional to
$1/\sqrt{f}$ and equals about 0.2 mrad/$\sqrt{Hz}$ at 1 Hz. This
corresponds to a bounded phase instability $\delta\mit \Phi$ of about
0.3 mrad. On the other hand, the input electronic noise floor is less than
-160 dBm/Hz, leading to an equivalent phase noise spectral density
lower than 0.3 mrad/$\sqrt{Hz}$ if the signal to be measured is above
1 pW (-90 dBm). For higher amplitude signals, the phase noise of PLLs
dominates and limits the resolution
\cite{sullivan1990characterization,durdaut2019equivalence}.

The above specifications are achieved by extremely careful design of
radio frequency circuit, power and clock distribution, feedback
temperature control of signal sources, as well as step-by-step
shielding to meet the stringent requirement of ElectroMagnetic
Compatibility (EMC) for over 80 dB channel isolation. Proper device selection is required. If the PLL is replaced by a
commercial microwave signal generator (e. g. R$\&$S SMB100A), the
phase noise is approximately -70 dBc/Hz at 1 Hz offset when the signal
frequency is 1 GHz, comparable with our results 0.2 mrad/$\sqrt{Hz}$ at 1 Hz offset frequency shown in Fig. 1(b). However, the phase noise spectral density
$S_\phi$(f) of our setup is proportional to 1/f down to 0.01 Hz, while the phase noise of typical microwave signal generator diverges at
frequency below 1 Hz. The rapid increase of phase noise as offset
frequency decreases suggests an unbounded drift of phase in hundreds
of second, while our setup remains stable during the three day period
of the data in Fig. 2(b). Our instrument focuses on long-term phase
detection of small signals at a single frequency. Our typical
resolution band width (RBW) is less than 0.3Hz, so that it can resolve
signals lower than -160 dBm. This is 4 orders of magnitude lower than
the about -120 dBm noise background of a commonly used vector network
analyzer (e.g., Agilent Technologies E5071C).  Furthermore, our
homemade instrument has the same low noise background compared to
commercial RF lock-in amplifiers (e.g., Zurich GHFLI), which implement
an all-digital demodulation scheme. In summary, by sacrificing the
flexibility, we achieve a very low noise background and extremely high
long-term phase stability \cite{wu2023morphing}.

The device used in this demonstration is fabricated on an undoped
bulk GaAs. We evaporate a 5-$\mu\text{m}$-period interdigital
transducer (IDT) on each side of sample. 50 $\Omega$ resistors are
connected in parallel to each IDT for broadband impedance
matching. The SAW is generated at the emitter IDT by the input voltage
$V_{\text{in}}$, propagates through the device and is captured by the
receiver IDT on the opposite side of the sample as a voltage output
$V_{\text{out}}$. We install the device in a dilution refrigerator and measure the attenuation and phase delay of $V_{\text{out}}$ using our lock-in setup. We can deduce the S-parameter $S_{21}$ with respect to $V_{\text{in}}$. Fig. 1(c) shows the $|S_{21}|$ and $\mit \Phi$
at different frequencies taken at 6.2 K. The $|S_{21}|$ peak indicates
the resonance condition when the stimulated SAW wavelength $\lambda$
equals the IDT period $\lambda_{\text{r}}$. The $\simeq$2 MHz
oscillations in $|S_{21}|$ are probably due to the echoes caused by the
finite reflectivity of the two IDTs. The SAW traveling time from the
emitter to the receiver IDTs can be measured by
$\tau=\frac{1}{2\pi}d\mit \Phi/df=\frac{L}{\lambda}\cdot
f_{\text{r}}^{-1}$, where $L$ is the nominal propagating
distance. $N=\frac{L}{\lambda}$ is an integer
defined by the lithography, i.e. the number of beats between the
emitter and receiver if the frequency is close to $f_{\text{r}}$. We
note that $\tau$ and $f_{\text{r}}$ remain almost unchanged when
temperature is below 6.5 K. Therefore, we can deduce the device
parameters at $T=0$ K to be $^0f_r=566.2$ MHz,
$^0v=\lambda_{\text{r}} ^0f_{\text{r}}=2830$ m/s, $^0\tau=1.1\mu$ s,
$L=3.1$ mm and $N=620$.

\begin{figure}[!htbp]
  \includegraphics[width=0.45\textwidth]{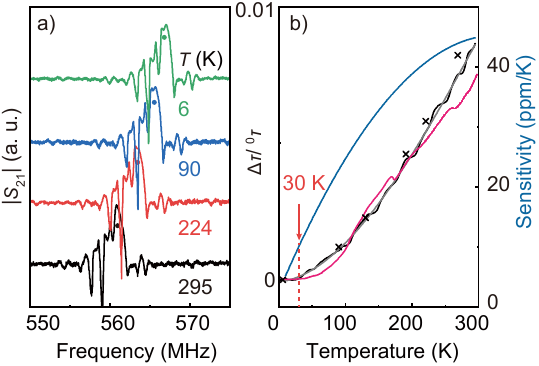}
  \caption{(a) $|S_{21}|$ of the device measured at
          different temperatures using Vector Network Analyzer. The
          traces are shifted vertically for clarity. (b) Black and pink curves present the temperature dependence of the propagation time shift $\Delta\tau/^0\tau$ measured by the phase approach (see text) using two different GaAs devices with 
          5 and 20$\mu$m IDT periods, respectively. The grey curve is the 3rd
          polynomial fitting of the 5$\mu$m period data. Black
          cross markers are deduced from the resonance frequency extracted from the data in Fig. 2(a) as verification for our method. Blue curve is the thermometer sensitivity $^0\tau^{-1}d\tau/dT$ derived from taking the derivative of the grey curve.}
       \label{Fig2}
\end{figure} 

Besides using the resonance frequency, we can significantly improve
the resolution of SAW velocity by directly measuring the propagation
time from the phase delay (the phase approach). We choose a frequency
$f\approx f_{\text{r}}$ so that the $|S_{21}|$ is close to its maximum. We then
measure the phase $-\pi<\mit \Phi<\pi$, and calculate the corresponding
delay time by $\tau=\frac{\mit \Phi}{2\pi f}+\frac{N}{f}$. Note that
$\mit \Phi$ = 0 and $\tau=N\cdot f^{-1}$ if $f=f_{\text{r}}$, consistent with the mode
where the SAW reflected from the receiver IDT constructively
interferes with the emitting wave. 

Our device can be used as a high resolution thermometer operating
from below 30 K to above 300 K. Fig. 2(a) shows the $|S_{21}|$ taken by
the Vector Network Analyzer at different temperatures, where the
resonance frequency $f_{\text{r}}$ decreases as the temperature increases. The
estimated thermal expansion $\Delta L/L$ is about two orders of magnitude smaller than the observed $\Delta f_{\text{r}}/f_{\text{r}}$, so that the SAW
velocity change dominates this temperature-dependent shift and
$\Delta f_{\text{r}}/^0f_{\text{r}}= \Delta v/^0v=-\Delta\tau/^0\tau$, where $^0f_{\text{r}}$,
$^0v$ and $^0\tau$ are the resonance frequency, SAW velocity and
propagation time at $T=0$ K. In Fig. 2(b), we summarize the
$\Delta \tau/^0\tau$ deduced from the resonance frequency shift
(symbols) and from our phase approach (curves). This
$\Delta\tau/^0\tau$ vs. $T$ curve can be fitted as
$\frac{\Delta\tau}{^0\tau} = (-0.8 + 2.6T + 1.3*10^{-1}T^2 -
1.3*10^{-4}T^3)$ \cite{powlowski2019temperature}. Using this curve as a calibration, we can derive the
temperature $T$ from the measured $\tau$. The sensitivity of the
SAW thermometer, $^0\tau^{-1}d\tau/dT$ reaches about 40 ppm/K at room temperature and decreases to 10 ppm/K when the temperature drops to 30 K. The shrinkage of sensitivity limits the minimum operating temperature of the SAW thermometer. 

The parameter $N$ is a well-defined integer by lithography and $^0v$ is an intrinsic parameter in high-purity materials such as GaAs. This allows us to
use one calibration curve for different devices. The phase instability
of our lock-in setup is sub-mrad, and the $\tau$ resolution is
$N$-times enhanced by the delay-line structure. Therefore we can
resolve $\tau$ as small as the $\frac{\delta\mit\Phi}{2\pi N}\sim $ 0.1ppm,
leading to mK-level resolution at room temperature. It is worth
mensioning that while a large $N$ leads to high resolution, it poses a
restriction on the measurement frequency used in the phase approach,
i.e. $f$ should be close to $f_{\text{r}}$ so that $N|f-f_{\text{r}}|/f_{\text{r}}<0.5$. An
additional phase $2\pi [N|f-f_{\text{r}}|/f_{\text{r}}]$ should be added/subtracted to
$\mit\Phi$, $[\cdot]$ means rounding algorithm, if the chosen $f$ is too
far away from the resonance frequency $f_{\text{r}}$. Properly choosing the
distance and period of the IDTs can compromise between the sensitivity
and the ease of use.

\begin{figure}[!htbp]
	\includegraphics[width=0.45\textwidth]{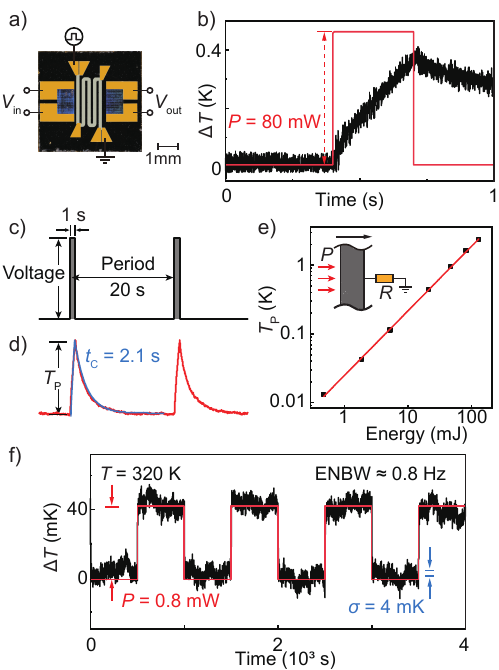}
	\caption{(a) The picture of our demonstration sensor device.
          (b) The time-domain response of temperature change using
          10kHz sampling rates. The red trace is the heating power. (c
          \& d)
          The pulsed heating voltage applied to the heater, and the
          measured temperature response of the SAW thermometer. The
          blue curve is the numerically simulated result. (e) Measured
          temperature peak $T_{\text{P}}$ as a function of
          heating energy per pulse. Inset is our model to
          describe the heating process. (f) Long-term measurement with
          a 1000-s-period, 0.8 mW amplitude square wave heating.}
       \label{Fig3}
\end{figure}

SAW sensors have an ultra-fast response to external environment, and
measuring phase delay allows us to read the sensor response with
minimal delay. In order to study the dynamic response of the SAW
thermometer, we evaporate a platinum resistance wire (666 $\Omega$) as
a heating element between the IDTs, see Fig. 3(a). The device is
installed in a vacuum thermostat where the temperature is kept
slightly above the room temperature at 320 K. We use $f=561.4$ MHz and
-40 dBm RF input power so that the SAW heating is negligible. In Fig. 3(b),
we apply an 80 mW, 0.3-second-wide square heating pulse and record the
output at 10 kHz sampling rate. From the Fig. 2(b) curve, the measured
phase $\mit\Phi$ is converted to temperature variation $\Delta T$ using a
sensitivity $(2\pi N)^{-1}d\mit\Phi/dT\simeq$ 46 ppm/K. In
Fig. 3(b), $\Delta T$ increases immediately
when the heater is switched on, and we observe no delay of more than a
few milliseconds.

We can describe the timeline of the thermal process for better
understanding. The temperature at the device surface increases
immediately when the current through the Pt wire generates heat. The
heat defuses into the GaAs substrate which reaches a thermally steady
state within a few milliseconds. The temperature sensor output
responds as soon as the SAW reaches the receiver IDT after a propagation delay of 1 $\mu$s. In order to investigate this process, we apply
20-s-separated 1-s-wide heating pulses with different power. A typical
measured response is shown in Fig. 3(d), and the peak height $T_{\text{P}}$ is summarized as a function of heating power for each pulse in Fig. 3(e).
Each temperature peak in Fig. 3(d) has an exponential decay time
$t_{\text{c}}\simeq 2.1$ s after each heating pulse. This decay time can be
explained by the model shown in the inset of Fig. 3(e). Here the
evolution of device temperature $T$ is described by the equation
$P = C_{\text{D}} \cdot dT/dt - T/R$, where $P$ is the heating power, $C_{\text{D}}$ is
the thermal capacitance of device and $R$ is the thermal resistance
between the sensor and the heat sink. By linearly fitting $T_{\text{P}}$ and the energy of each pulse in Fig. 3(e), we can deduce the thermal capacitance of
our device to be $C_{\text{D}} = 53$ mJ/K. A more accurate estimate of $C_{\text{D}}$ is 42
mJ/K once the finite thermal resistance $R$ is taken in account. $R$ is
estimated from the decay time as $R=t_{\text{c}}/C_{\text{D}}\simeq$ 50 mK/mW. The
numerical simulation is shown by the blue curve in Fig. 3(d). 

We can apply a constant heating power so that the device will reach a
steady state determined by the finite $R$. In Fig. 3(f), we apply a 745
mV, 1000-second-period voltage to the heater, generating a heating power of 0.8 mW. We observe about 41 mK temperature increase of the
sample when the heater is on, corresponding to a thermal resistance of $R=51$ mK/mW. This is consistent with the results of the pulsed heating measurement in Fig.
3(c-e). The standard deviation of $\Delta T$ in Fig. 3(f) is only 4 mK
at 0.8 Hz measurement bandwidth, evidencing the exceptionally high
performance of this proposed temperature sensor.

In conclusion, we introduce an approach to measure the propagation
time of the SAW delay-line devices with 0.1-ppm resolution, and
demonstrate that these devices show considerable promise for
high-sensitivity and high-stability temperature sensors. This technique
would be of interest for future applications due to its accuracy,
sensitivity and fast response time.

\begin{acknowledgments}
The work is supported by the National Key Research and Development Program of China (Grant No. 2021YFA1401900 and 2019YFA0308403) and the National Natural Science Foundation of China (Grant No. 92065104 and 12074010).
\end{acknowledgments}

\bibliographystyle{apsrev4-1}
\bibliography{bib_full_2}

\end{document}